\documentclass[a4paper,12pt]{article}
\usepackage{amsmath, amsthm, amsfonts, amssymb}
\usepackage{graphicx}
\usepackage[mathscr]{eucal}

\theoremstyle{plain}
\newtheorem*{theorem}{Theorem}

\newtheorem*{lemma}{Lemma}

\def\e{\varepsilon}

\def\l{\lambda}
\def\p{\partial}

\def\t{\widetilde}

\def\Odr{\mathcal{O}}
\def\H{W_2}
\def\Hloc{W_{2,loc}}

\def\di{\,\mathrm{d}}

\def\la{\langle}
\def\ra{\rangle}

\DeclareMathOperator{\RE}{Re}

\DeclareMathOperator{\supp}{supp}

\begin{document}

\allowdisplaybreaks

\begin{center}

\textbf{\Large On spectrum of a Schr\"odinger operator with a
fast oscillating compactly supported potential}

\bigskip
\bigskip

{\large D.I. Borisov$^\natural$, R.R.
Gadyl'shin$^\sharp$}\footnotetext[1]{The work is partiall
supoorted by RFBR (05-01-97912-r\_agidel) and by the programs
''Leading scientific schools'' (NSh-1446.2003.1) and
''Universities of Russia'' (UR.04.01.484).}

\begin{quote}
\emph{Bashkir State Pedagogical University, October rev. st.,
3a,  Ufa, Russia, 450000. \\ $^\natural$E-mail:
\texttt{borisovdi@yandex.ru}, URL:
\texttt{http://borisovdi.narod.ru/} }
$^\sharp$\emph{E-mail:
\texttt{gadylshin@yandex.ru}, URL:
\texttt{http://gadylshin.narod.ru/} }
\end{quote}

\end{center}

\begin{abstract}
We study the phenomenon of an eigenvalue emerging from essential
spectrum of a Schr\"odinger operator perturbed by a fast
oscillating compactly supported potential. We prove the
sufficient conditions for the existence and absence of such
eigenvalue. If exists, we obtain the leading term of its
asymptotics expansion.
\end{abstract}

The present work is devoted to the study of the spectrum of the
operator
\begin{equation*}
H_\e:=-\frac{d^2}{dx^2}+V\left(x,\frac{x}{\e}\right)
\end{equation*}
in $L_2(\mathbb{R})$ with domain $\H^2(\mathbb{R})$. Here $\e$
is a small positive parameter, $V(x,\xi)$ is a complex-valued
$1$-periodic on $\xi$ function belonging to
$C^\infty(\mathbb{R}^2)$, such that for all  $\xi\in\mathbb{R}$
the support $\supp V(\cdot,\xi)$ is bounded uniformly on
 $\xi$.

The operator $H_0:=-\frac{\displaystyle d^2}{\displaystyle
dx^2}$ in $L_2(\mathbb{R})$ with domain $\H^2(\mathbb{R})$ is
self-adjoint, its discrete spectrum is empty while the essential
one coincides with the semi-axis $[0,+\infty)$. The
multiplication operator by the function
$V\left(x,\frac{x}{\e}\right)$ is $H_0$-compact. This is why by
Theorems~1.1,~5.35 in Chapter~I\!V of \cite{K} the operator
$H_\e$ is closed for all $\e$ and the essential spectra of the
operator $H_\e$ and $H_0$ are same.

The aim of the work is to study the existence and the
asymptotics behaviour of the eigenvalues of the operator $H_\e$,
tending to zero as $\e\to0$, in the case the mean value of the
function $V(x,\cdot)$ over period is zero  for all
$x\in\mathbb{R}$. Such eigenvalues can be also regarded as
emerging from the border of the essential spectrum when
perturbing the operator $H_0$ by the potential
$V\left(x,\frac{x}{\e}\right)$. We note that the phenomenon of
the eigenvalues emerging from the border of a essential spectrum
under the perturbation by the potential was treated in
\cite{Si}--\cite{KS} for the potentials of the form $\e U(x)$,
where $U(x)$ is a sufficiently rapidly decaying real potential.
This phenomenon was also considered in  \cite{G1} for the
perturbation $\e L_\e$, where $L_\e: \Hloc^2(\mathbb{R})\to
L_2(\mathbb{R};Q)$ is an arbitrary operator obeying a uniform on
$\e$ inequality $\|L_\e u\|_{L_2(\mathbb{R})}\leqslant C
\|u\|_{\H^2(Q)}$ for a finite interval $Q$ in the axis, and
$L_2(\mathbb{R};Q)$ is a subset of the functions from
$L_2(\mathbb{R})$ whose supports are in $\overline{Q}$. In
\cite{Si}--\cite{G1} for the mentioned perturbations the
existence of the eigenvalues emerging from the border of the
essential spectrum was studied. If exists, the leading term of
the asymptotics expansions for an eigenvalue was constructed.
Clearly, the perturbation considered in the articles cited do
not include the potential $V\left(x,\frac{x}{\e}\right)$.
Moreover, the perturbation described by this potential is not
regular in the sense that the multiplication operator by this
potential does not tend to zero uniformly as $\e\to0$.

Let a segment $M=[x_0,x_1]$ be such that for all
$\xi\in\mathbb{R}$ the inclusion $\supp V(\cdot,\xi)\subseteq M$
holds true. By $W$ we denote the set of $1$-periodic on  $\xi$
functions $u(x,\xi)$ from $C^\infty(\mathbb{R}^2)$ such that
$\supp u(\cdot,\xi)\subseteq M$ for all $\xi\in \mathbb{R}$. For
a function from $W$ its mean value over period will be indicated
as
\begin{equation*}
\la u(x,\cdot)\ra:=\int\limits_0^1 u(x,\xi)\di\xi.
\end{equation*}

\begin{lemma}\label{lm}
For each function $u\in W$ and each number $n\geqslant 1$ the
equality
\begin{equation*}
\int\limits_\mathbb{R} u\left(x,\frac{x}{\e}\right)\di x=
\int\limits_\mathbb{R} \la u(x,\cdot) \ra\di x+\Odr(\e^n)
\end{equation*}
holds true.
\end{lemma}

\begin{proof}
It is sufficient to prove the statement of the lemma in the case
$\la u(x,\cdot)\ra\equiv0$, since the general case is reduced to
this one by the change $\t u(x,\xi)=u(x,\xi)-\la u(x,\cdot)\ra$.
Let $\la u(x,\cdot)\ra\equiv0$. We denote
\begin{gather}
P[u](x,\xi):=\int\limits_0^\xi u(x,\tau)\di\tau+ \int\limits_0^1
\tau u(x,\tau)\di\tau.\nonumber
\end{gather}
Then $P[u]\in W$ and $\la P[u](x,\cdot)\ra=0$. Bearing in mind
an obvious equality
\begin{equation*}
u\left(x,\frac{x}{\e}\right)=\e\frac{d}{dx}P[u]
\left(x,\frac{x}{\e}\right)-\e\frac{\p P[u]}{\p
x}\left(x,\frac{x}{\e}\right),
\end{equation*}
we get
\begin{equation}\label{2}
\int\limits_\mathbb{R} u\left(x,\frac{x}{\e}\right)\di x=
-\e\int\limits_\mathbb{R}\frac{\p P[u]}{\p
x}\left(x,\frac{x}{\e}\right)\di x.
\end{equation}
Since $\frac{\displaystyle\p P[u]}{\displaystyle\p x}\in W$ и
$\left\la \frac{\displaystyle\p P[u]}{\displaystyle\p
x}(x,\cdot) \right\ra\equiv0$, it follows that the equality
(\ref{2}) is applicable to the function $\frac{\displaystyle\p
P[u]}{\displaystyle\p x}$ as well. Applying this equality as
many times as needed, we arrive at the statement of the lemma.
\end{proof}

We denote
\begin{gather}
k_2:=\frac{1}{2}\int\limits_\mathbb{R} \left\la
\big(P[V](x,\cdot)\big)^2\right\ra\di x.\label{10}
\end{gather}

The main result of the work is the following
\begin{theorem}\label{th}
Let
\begin{equation}\label{1.1}
\la V(x,\cdot) \ra\equiv0.
\end{equation}
Then
\begin{enumerate}
\def\theenumi{\arabic{enumi}}
\item
If $\RE k_2>0$, then there exists the unique eigenvalue $\l_\e$
of the operator $H_\e$, tending to zero as $\e\to0$. This
eigenvalue is simple and its asymptotics is of the form:
\begin{equation}\label{1}
\l_\e=-\e^4 k_2^2+\Odr(\e^5).
\end{equation}

\item
If $\RE k_2<0$, then the operator $H_\e$ has no eigenvalues
tending to zero as $\e\to0$.
\end{enumerate}
\end{theorem}

\begin{proof}[Доказательство]
It is easy to check that the function
\begin{equation*}
v(x,\xi):=\int\limits_0^\xi (\xi-\tau) V(x,\tau)\di\tau+\xi
\int\limits_0^1\tau V(x,\tau)\di\tau+v_1(x)
\end{equation*}
is a solution to the equation
\begin{equation}\label{3-}
\frac{\p^2 v}{\p\xi^2}=V,
\end{equation}
moreover,
\begin{equation}\label{3}
\frac{\p v}{\p\xi}=P[V].
\end{equation}
By  (\ref{1.1}) and the belonging $V\in W$ the function $v$ is
$1$-periodic on $\xi$, and, therefore, is bounded uniformly on
$(x,\xi)\in\mathbb{R}^2$. We assume that $v_1\in
C^\infty(\mathbb{R})$ and $\supp v_1\subseteq M$; then $v\in W$.

Let $q_\e(x,\xi):=1+\e^2 v(x,\xi)$. The multiplication operator
by the function $\t q_\e(x):=q_\e\left(x,\frac{x}{\e}\right)$
(we denote is by $Q_\e$) maps $L_2(\mathbb{R})$ in a one-to-one
way onto itself. This is why the eigenvalues of the operator
$H_\e$ coincide with those of the operator $Q_\e^{-1}H_\e Q_\e$.
By (\ref{3-}) and the definition of  $\t q_\e(x)$ the
representation $Q_\e^{-1}H_\e Q_\e=H_0-\e L_\e$ is valid, where
\begin{equation}\label{L}
L_\e=\e\frac{2} {\t q_\e(x)}\frac{d}{dx}
v\left(x,\frac{x}{\e}\right)
\frac{d}{dx}-\frac{f_\e\left(x,\frac{x}{\e}\right)} {\t
q_\e(x)},\quad f_\e=\e V v-\e\frac{\p^2 v}{\p x^2}-2\frac{\p^2
v}{\p x\p\xi}.
\end{equation}
Since $V,\,v\in W$, it follows that the supports of the
coefficients of the operator $L_\e$ lie in $M$ for all values of
$\e$, and the operator $L_\e: \Hloc^2(\mathbb{R})\to
L_2(\mathbb{R};M)$ meets a uniform on $\e$
inequality
\begin{equation}\label{Old}
\|L_\e u\|_{L_2(\mathbb{R})}\leqslant C\|u\|_{\H^2(M)}.
\end{equation}

We denote
\begin{equation}\label{4}
m_\e^{(1)}:=\int\limits_\mathbb{R} L_\e[1]\di x,\quad
m_\e^{(2)}:=\int\limits_\mathbb{R}
L_\e\left[\int\limits_\mathbb{R} |x-t| L_\e[1]\di t\right]\di
x,\quad k_\e:=\frac{\e}{2}m_\e^{(1)}+\frac{\e^2}{2} m_\e^{(2)}.
\end{equation}
The estimate (\ref{Old}) begin valid for the operator $L_\e$,
Theorem~1 of the work \cite{G1} implies that if
\begin{equation}\label{5}
k_\e=\e c_1+\e^2 c_2+\Odr(\e^3),\quad c_1,c_2=\mathrm{const},
\end{equation}
then the sufficient condition the operator $(H_0-\e L_\e)$ to
have the eigenvalue tending to zero as $\e\to0$ is the
inequality $\RE(c_1+\e c_2)>0$, while the sufficient condition
of the absence is the inequality $\RE(c_1+\e c_2)<0$. If $\RE(
c_1+\e c_2)>0$, then the operator $(H_0-\e L_\e)$ has the unique
eigenvalue tending to zero, this eigenvalue is simple and the
equality $\l_\e=-(\e c_1+\e^2 c_2)^2+\Odr(c_1\e^4+\e^5)$ holds
true.

Thus, in order to prove the theorem it is sufficient to
establish the equality (\ref{5}) с $c_1=0$, $c_2=k_2$. Let us
prove it. For the sake of simplicity of calculations we set
\begin{equation*}
v_1(x):=\frac{1}{2}\int\limits_0^1(\tau-\tau^2)V(x,\tau)\di\tau.
\end{equation*}
It is easy to see that in this case
\begin{equation}\label{3--}
\la v(x,\cdot) \ra=0,
\end{equation}
and, therefore,
\begin{equation}\label{3+}
\left\la \frac{\p^2 v}{\p x^2}(x,\cdot)\right\ra= \left\la
\frac{\p^2 v}{\p x\p\xi}(x,\cdot)\right\ra=0.
\end{equation}
We denote $\t f_\e(x):=f_\e\left(x,\frac{x}{\e}\right)$. From
(\ref{4}), (\ref{L}) and the definition of $\t q_\e$ by Lemma
and the equalities (\ref{3+}), (\ref{3-}), (\ref{3}) it follows
that
\begin{equation}\label{7a}
\begin{aligned}
\frac{\e m_\e^{(1)}}{2}&=-\frac{\e}{2}\int\limits_\mathbb{R}
\frac{\t f_\e(x)} {\t q_\e(x)}\di
x=-\frac{\e^2}{2}\int\limits_\mathbb{R}
V\left(x,\frac{x}{\e}\right) v\left(x,\frac{x}{\e}\right)\di
x+\Odr(\e^3)
\\
&=-\frac{\e^2}{2}\int\limits_\mathbb{R} \la V(x,\cdot)
v(x,\cdot)\ra \di
x+\Odr(\e^3)=-\frac{\e^2}{2}\int\limits_\mathbb{R} \left\la
v(x,\cdot) \frac{\p^2 v}{\p\xi^2}(x,\cdot)\right\ra \di
x+\Odr(\e^3)\\ &=\frac{\e^2}{2}\int\limits_\mathbb{R} \left\la
\left(\frac{\p v}{\p\xi}(x,\cdot)\right)^2\right\ra \di
x+\Odr(\e^3)=\e^2 k_2+\Odr(\e^3),
\end{aligned}
\end{equation}
where $k_2$ is from (\ref{10}).

Since by the definition of $\t q_\e$
\begin{equation*}
\frac{d}{dx}\ln \t q_\e(x)=\frac{\e^2} {\t q_\e(x)}\frac{d}{dx}
v\left(x,\frac{x}{\e}\right),
\end{equation*}
integrating by parts and taking into account that for each
function $g\in C_0(\mathbb{R})$ the equality
\begin{align*}
\frac{d^2}{dx^2} \int\limits_\mathbb{R}|x-t| g(t)\di t=2g(x)
\end{align*}
is valid, from  (\ref{4}), (\ref{L}) we obtain:
\begin{align*}
\frac{\e^2 m_\e^{(2)}}{2}&=-\e\int\limits_\mathbb{R} \ln \t
q_\e(x)\frac{d^2}{dx^2} \int\limits_\mathbb{R}|x-t|\frac{ \t
f_\e(t)}{ \t q_\e(t)}\di t\di x +\frac{\e^2}{2}
\int\limits_{M^2} \frac{|x-t| \t f_\e(x) \t f_\e(t)}{\t q_\e(x)
\t q_\e(t)}\di t\di x
\\
&=-2\e\int\limits_\mathbb{R} \frac{\t f_\e(x)\ln\t q_\e(x)} {\t
q_\e(x)}\di x+ \e^2\int\limits_{x_0}^{x_1}\frac{\t f_\e (x)} {\t
q_\e(x)} \left(\int\limits_{x_0}^x \frac{(x-t)\t f_\e(t)} {\t
q_\e(t)}\di t\right)\di x.
\end{align*}
The first integral in the right hand side of the equality
obtained is of order $\Odr(\e^3)$. Integrating by parts in the
second integral we deduce:
\begin{align*}
&\int\limits_{x_0}^{x_1}\frac{\t f_\e(x)} {\t q_\e(x)}
\left(\int\limits_{x_0}^x \frac{(x-t)\t f_\e(t)} {\t q_\e(t)}\di
t\right)\di x = \int\limits_{x_0}^{x_1} x \t f_\e(x)
\left(\int\limits_{x_0}^x \t f_\e(t)\di t\right)\di x-
\\
&-\int\limits_{x_0}^{x_1}\t f_\e(x) \left(\int\limits_{x_0}^x t
\t f_\e(t)\di t\right) \di x+\Odr(\e^2)=2\int\limits_{x_0}^{x_1}
x \t f_\e(x) \left(\int\limits_{x_0}^x \t f_\e(t)\di t\right)\di
x-
\\
&-\int\limits_\mathbb{R} x \t f_\e(x)\di x\int\limits_\mathbb{R}
\t f_\e(x)\di x+\Odr(\e^2)=-\int\limits_{x_0}^{x_1}
\left(\int\limits_{x_0}^x \t f_\e(t)\di t\right)^2 \di x+
\\
&+\int\limits_\mathbb{R} (x_1-x)\t f_\e(x)\di
x\int\limits_\mathbb{R} \t f_\e(x)\di x+\Odr(\e^2).
\end{align*}
It follows from the definition of the function $\t f_\e$, the
formula for $f_\e$ from (\ref{L}), the equalities (\ref{3+}) and
Lemma that the second term in the right hand side of the last
equality is of order $\Odr(\e^2)$. Taking into account the
definition of the function $f_\e$ once again as well as the
equality
\begin{equation*}
\frac{\p^2 v }{\p x\p\xi}\left(x,\frac{x}{\e}\right)=
\e\frac{d}{dx}\frac{\p v}{\p x}\left(x,\frac{x}{\e}\right)
-\e\frac{\p^2 v}{\p x^2}\left(x,\frac{x}{\e}\right),
\end{equation*}
we obtain
\begin{equation*}
-\int\limits_{x_0}^{x_1} \left(\int\limits_{x_0}^x \t f_\e(t)\di
t\right)^2 \di x=-4\int\limits_M \left(\int\limits_{x_0}^x
\frac{\p^2 v }{\p x\p\xi}\left(t,\frac{t}{\e}\right)\di t
\right)^2 \mathrm{d} x+\Odr(\e^2)=\Odr(\e^2).
\end{equation*}
Thus,
\begin{equation*} \int\limits_{x_0}^{x_1}\frac{\t f_\e(x)}
{\t q_\e(x)} \left(\int\limits_{x_0}^x \frac{(x-t)\t f_\e(t)}
{\t q_\e(t)}\di t\right)\di x=\Odr(\e^2),
\end{equation*}
hence, $\e^2 m_\e^{(2)}=\Odr(\e^4)$, and the equality (\ref{5})
с $c_1=0$, $c_2=k_2$ now follows from (\ref{4}) and (\ref{7a}).
\end{proof}

It also follows from Theorem~1 of the paper \cite{G1} that all
the eigenvalues of the operator $H_\e$ except the one tending to
zero (if they exist) must tend to infinity as $\e\to0$. If $V$
is a non-zero real-valued function, then all the eigenvalues of
the operator $H_\e$ are real and negative. Moreover, in this
case $k_2$ is real and positive, hence, the operator $H_\e$ has
the unique eigenvalue, this eigenvalue is simple, tends to zero
and has the asymptotics (\ref{1}).

\end{document}